\documentclass[onecolumn]{IEEEtran}
\usepackage[latin9]{inputenc}
\usepackage{verbatim}
\usepackage{amsmath}
\usepackage{amsthm}
\usepackage{amssymb}
\usepackage{graphicx}
\usepackage[bookmarks=false]{hyperref}
\makeatletter
\theoremstyle{plain}
\newtheorem{thm}{\protect\theoremname}
\theoremstyle{definition}
\newtheorem{problem}[thm]{\protect\problemname}
\theoremstyle{plain}
\newtheorem{lem}[thm]{\protect\lemmaname}

\IEEEoverridecommandlockouts
\usepackage{cite}
\usepackage{amsfonts}%\usepackage{algorithmic}
\usepackage{textcomp}
\def\BibTeX{{\rm B\kern-.05em{\sc i\kern-.025em b}\kern-.08em
    T\kern-.1667em\lower.7ex\hbox{E}\kern-.125emX}}

\providecommand{\lemmaname}{Lemma}
\providecommand{\problemname}{Problem}
\providecommand{\theoremname}{Theorem}

\makeatother

\providecommand{\lemmaname}{Lemma}
\providecommand{\problemname}{Problem}
\providecommand{\theoremname}{Theorem}

\begin{document}

\title{On the detection of low rank matrices in the high-dimensional regime.}

\author{Antoine~Chevreuil and ~Philippe~Loubaton\\
	\IEEEauthorblockA{Gaspard Monge Computer Science Laboratory (LIGM) - UMR 8049 CNRS \protect\\		
				 Université de 
				Paris-Est/Marne-la-Vallée \protect\\
				5 Bd. Descartes 77454 Marne-la-Vallée (France)}
	}

%\author{Antoine~Chevreuil and ~Philippe~Loubaton% <-this % stops a space
%	\IEEEcompsocitemizethanks{\IEEEcompsocthanksitem 
%	Gaspard-Monge Computer Science Laboratory (LIGM) - UMR 8049 CNRS \protect\\		
%		 Université de 
%		Paris-Est/Marne-la-Vallée UMR 8049\protect\\
%		5 Bd. Descartes 77454 Marne-la-Vallée (France).\protect\\
%		% note need leading \protect in front of \\ to get a newline within \thanks as
%		% \\ is fragile and will error, could use \hfil\break instead.
%		E-mail: $\{$chevreui,loubaton$\}$@u-pem.fr
%	%\thanks{blabla}
%	}}

\maketitle
\begin{abstract}
We address the detection of a  low rank $n\times n$ matrix $\mathbf{X}_0$ from the noisy observation 
${\bf X}_0 + {\bf Z}$ when $n\to \infty$, where ${\bf Z}$ is a complex Gaussian random matrix with 
independent identically distributed $\mathcal{N}_c(0,\frac{1}{n})$ entries.
Thanks to large random matrix theory results, it is now well-known that if the largest singular value  $\lambda_1(\mathbf{X}_0)$ of ${\bf X}_0$ verifies $\lambda_1(\mathbf{X}_0)>1$, then it is possible to exhibit consistent tests. In this contribution, we prove {\it a contrario  } that under the condition $\lambda_1(\mathbf{X}_0)<1$, there are no consistent tests. Our proof is inspired by previous works devoted to the case of rank 1 matrices ${\bf X}_0$.  
\end{abstract}


\section{Introduction}

\label{sec:intro}

The problem of testing whether an observed $n_{1}\times n_{2}$ matrix
${\bf Y}$ is either a zero-mean independent identically distributed
Gaussian random matrix ${\bf Z}$ with variance $\frac{1}{n_{2}}$,
or ${\bf X}_{0}+{\bf Z}$ for some low rank deterministic matrix ${\bf X}_0$,
called also a \textit{spike}, is a fundamental problem arising in
numerous applications such as the detection of low-rank multivariate
signals or the Gaussian hidden clique problem. When the two dimensions
$n_{1},n_{2}$ converge towards $\infty$ in such a way that $n_{1}/n_{2}\rightarrow c>0$
(the rank of ${\bf X}_{0}$ remaining fixed), known results on the
so-called additive spiked large random matrix models have enabled  to re-consider
this fundamental detection problem (see {\it e.g.  } \cite{nadakuditi-edelman-2008},
\cite{bianchi-et-al-2011}, \cite{benaych2012singular}). It was established
a long time ago (see {\it e.g. } \cite{bai-silverstein-book} and the references
therein) that in the above asymptotic regime, the largest singular
value $\lambda_{1}({\bf Z})$ of ${\bf Z}$ converges almost surely
towards $1+\sqrt{c}$. More recently, under mild technical extra assumptions,
\cite{benaych2012singular} proved that $\lambda_{1}({\bf X}_{0}+{\bf Z})$
still converges towards $1+\sqrt{c}$ if $\lambda_{1}({\bf X}_{0})$
converges towards a limit strictly less than $c^{1/4}$. On the contrary, 
if the limit of $\lambda_{1}({\bf X}_{0})$ is strictly greater than
$c^{1/4}$, then $\lambda_{1}({\bf X}_{0}+{\bf Z})$ converges towards
a limit strictly greater than $1+\sqrt{c}$. This result implies that
the Generalized Likelihood Ratio Test (GLRT) is consistent ({\it i.e. } both
the probability of false alarm and the probability of missed detection
converge towards $0$ in the above asymptotic regime) if and only if $\lambda_{1}(\mathbf{X}_{0})$
is above the threshold $c^{1/4}$. In order to simplify the exposition,
we assume from now on that $n_{1}=n_{2}=n$, so that ratio $c$ reduces
to $1$.

While the detection problem was extensively addressed in the zone $\lambda_1({\bf X}_0) > 1$, 
the case where  $\lambda_1({\bf X}_0) < 1$ was much less studied. 
Montanari \emph{et al. }\cite{montanari-reichman-zeitouni-2017}
consider  the zone  $\lambda_{1}(\mathbf{X}_{0})<1$ when
${\bf X}_{0}$ is a rank 1 matrix. Thanks to information geometry
tools, \cite{montanari-reichman-zeitouni-2017} prove that, in this region, it is impossible
to find a consistent test for the detection of the spike. Irrespective
of the standard random matrix tools, this approach is extended 
in \cite{montanari-reichman-zeitouni-2017} to the more general case when ${\bf X}_{0}$ and ${\bf Z}$
are tensors of order $d\geq3$; namely, if the Frobenius norm of the
tensor $\mathbf{X}_{0}$ is stricly less than a threshold depending
in $d$, then the probability distributions of the observation under
the two hypotheses are asymptotically undistinguishable, so that any
detection test cannot behave better than a random guess. This property,
which is stronger than the non-existence of a consistent test, does
not hold in the matrix case $d=2$: see for instance \cite{onatski-moreira-hallin-2013}
where a non-consistent test is exhibited that has a better performance
than a random guess. 
When the spike follows a probabilistic model, the replica method gives an information-theoretic threshold for the estimation problem: see  \cite{Lelarge-Miolane:2016} and the references therein.  A connection with spectral methods is provided in section 2.3 of \cite{Lelarge-Miolane:2016}.   
In this paper, we focus on the case where $\mathbf{X}_{0}$
has general rank $r$.
Our contribution is to prove that under $\lambda_{1}(\mathbf{X}_{0})<1,$ the
consistent detection is impossible. While this theoretical result
is not unexpected, we believe that it provides a better understanding
of the above fundamental detection problem in large dimensions without resorting
to the machinery of large random matrices.

\section{Model, notation, asumption \label{sec:Model}}

The set of complex-valued matrices $\mathbb{C}^{n\times n}$ is a
complex vector-space endowed with the standard scalar product $\left\langle \mathbf{X},\mathbf{Y}\right\rangle =\text{Tr}(\mathbf{X}\mathbf{Y}^{*})$
and the Frobenius norm $\left\Vert \mathbf{X}\right\Vert _{F}=\sqrt{\left\langle \mathbf{X},\mathbf{X}\right\rangle }.$ The spectral norm 
of a matrix ${\bf X}$ is denoted by $\| {\bf X} \|_{\scriptscriptstyle 2}$. 
The spike (``the signal'') is assumed to be a matrix of fixed rank
$r$ and hence admits a SVD such as 
\begin{equation}
\mathbf{X}_{0}=\sum_{j=1}^{r}\lambda_{j}\mathbf{u}_{j}\mathbf{v}_{j}^{*}={\bf U}\boldsymbol{\Lambda}{\bf V}^{*}\label{eq:SVD X0}
\end{equation}
where $\lambda_{j}=\lambda_{j}(\mathbf{X}_{0})$ are the singular
values of $\mathbf{X}_{0}$ sorted in descending order and where $\boldsymbol{\Lambda}$
is the diagonal matrix gathering the $(\lambda_{j})_{j=1,\ldots,r}$
in the descending order. As $\mathbf{X}_0$ has to be defined for any
$n$, we impose a non-erratic behavior of $\mathbf{X}_{0}$, namely
that all its singular values $(\lambda_{j})_{j=1, \ldots, r}$ do not depend on $n$ for $n$ large enough.
This hypothesis could be replaced by the condition that $(\lambda_{j})_{j=1,\ldots,r}$
all converge towards a finite limit at an \emph{ad'hoc} rate. However,
this would introduce purely technical difficulties.

The noise matrix $\mathbf{Z}$ is assumed to have i.i.d. entries distributed
as $\mathcal{N}_{c}(0,1/n)$. We consider the alternative
$\mathcal{H}_{0}:\ \mathbf{Y}=\mathbf{Z}$ versus $\mathcal{H}_{1}:\ \mathbf{Y}=\mathbf{X}_{0}+\mathbf{Z}.$
We denote by $p_{1,n}({\bf y})$ the probability probability density
of $\mathbf{Y}$ under $\mathcal{H}_{1}$ and $p_{0,n}({\bf y})$
the density of $\mathbf{Y}$ under $\mathcal{H}_{0}$. $\mathcal{L}(\mathbf{Y})=\frac{p_{1,n}(\mathbf{Y})}{p_{0,n}(\mathbf{Y})}$
is the likelihood ratio and we denote by $\mathbb{E}_{0}$ the expectation
under $\mathcal{H}_{0}$. We now recall the fundamental information
geometry results used in \cite{montanari-reichman-zeitouni-2017}
in order to address the detection problem.The following property
is well known (see also \cite{banks-vershynin2017} section 3): 
 if $\mathbb{E}_{0}\left[\mathcal{L}(\mathbf{Y})^{2}\right]$ is
bounded, then no consistent detection test exists. 
We however mention that this is a sufficient conditions:
 $\mathbb{E}_{0}\left[\mathcal{L}(\mathbf{Y})^{2}\right]$
unbounded does not imply the existence of consistent tests.

\section{Expression of the second-order moment. }

The density of $\mathbf{Z}$, seen as a collection of $n^{2}$ complex-valued
random variables, is  $p_{0,n}(\mathbf{z})=\left(\frac{n}{\pi}\right)^{n^{2}} \exp\left(-n\left\Vert \mathbf{z}\right\Vert _{F}^{2}\right)$. On the one
hand, we notice that the study of the second-order moment of the likelihood ratio is not suited
to the deterministic model of the spike as presented previously. Indeed,
in this case $\mathbb{E}_{0}\left[\mathcal{L}(\mathbf{Y})^{2}\right]$has
the simple expression $\exp\left(2n\left\Vert \mathbf{X}_{0}\right\Vert _{F}^{2}\right)$
and always diverges. On the other hand, the noise matrix shows an
invariance property: if $\boldsymbol{\Theta}_{1},\boldsymbol{\Theta}_{2}$
are unitary $n\times n$ matrices , then the density of $\boldsymbol{\Theta}_{1}\mathbf{Z}\boldsymbol{\Theta}_{2}$
equals this of $\mathbf{Z}$. We hence modify the data according to
the procedure: we pick two independent unitary $\boldsymbol{\Theta}_{1},\boldsymbol{\Theta}_{2}$
according to the Haar measure (which corresponds to the uniform distribution
on the set of all unitary $n\times n$ matrices), and change the data
tensor $\mathbf{Y}$ according to $\boldsymbol{\Theta}_{1}\mathbf{Y}\boldsymbol{\Theta}_{2}.$
As said above, this does not affect the distribution of the noise,
but this amounts to assume a certain prior on the spike. Indeed, this
amounts to replace $\mathbf{u}_{i}$ by $\boldsymbol{\Theta}_{1}\mathbf{u}_{i}$
and $\mathbf{v}_{i}$ by $\boldsymbol{\Theta}_{2}^{*}\mathbf{v}_{i}$.
In the following, the data and the noise tensors after this procedure
are still denoted respectively by $\mathbf{Y}$ and $\mathbf{Z}$.

We are now in position to give a closed-form expression of the second-order
moment of $\mathcal{L}({\bf Y})$ . We have $p_{1,n}(\mathbf{Y})=\mathbb{E}_{X}\left[p_{0,n}(\mathbf{Y}-\mathbf{X})\right]$
where $\mathbb{E}_{X}$ is the mathematical expectation over the prior
distribution of the spike, or equivalently over the Haar matrices
$\boldsymbol{\Theta}_{1},\boldsymbol{\Theta}_{2}$. It holds that
$\mathbb{E}_{0}\left[\mathcal{L}(\mathbf{Y})^{2}\right]=\mathbb{E}\left[\exp\left(2n\mathfrak{R}\left\langle \mathbf{X},\mathbf{X}'\right\rangle \right)\right]$
where the expectation is over independent copies $\mathbf{X},\mathbf{X}'$
of the spike ($\mathfrak{R}$ stands for the real part); $\mathbf{X}$
and $\mathbf{X}'$ being respectively associated with $(\boldsymbol{\Theta}_{1},\boldsymbol{\Theta}_{2})$
and $(\boldsymbol{\Theta}_{1}',\boldsymbol{\Theta}_{2}')$, $\mathbb{E}_{0}\left[\mathcal{L}(\mathbf{Y})^{2}\right]$
has the expression 
\[
\mathbb{E}\left[\exp\left(2n\mathfrak{R}\text{Tr}\left(\boldsymbol{\Theta}_{1}{\bf X}_{0}\boldsymbol{\Theta}_{2}\left(\boldsymbol{\Theta}'_{2}\right)^{*}{\bf X}_{0}^{*}\left(\boldsymbol{\Theta}'_{1}\right)^{*}\right)\right)\right].
\]
As $\mathbf{\Theta}_{k}$ and $\mathbf{\Theta}_{k}'$ are Haar and
independent, then $\left(\mathbf{\Theta}'_{1}\right)^{*}\mathbf{\Theta}_{1}$
and $\mathbf{\Theta}_{2}\left(\mathbf{\Theta}'_{2}\right)^{*}$ are
also independent, Haar distributed and it holds 
\begin{align}
\mathbb{E}_{0}\left[\mathcal{L}(\mathbf{Y})^{2}\right] & =\mathbb{E}\left[\exp\left(2n\eta\right)\right],\label{eq:second order moment}
\end{align}
where the expectation is over the independent Haar matrices $\mathbf{\Theta}_{1},\mathbf{\Theta}_{2}$
and $\eta=\mathfrak{R}\text{Tr}\left(\boldsymbol{\Theta}_{1}\mathbf{X}_{0}\boldsymbol{\Theta}_{2}\mathbf{X}_{0}^{*}\right)$.
The ultimate simplification comes from the decomposition \eqref{eq:SVD X0}
which implies that 
\begin{equation}
\eta=\mathfrak{R}\text{Tr}\left(\mathbf{\boldsymbol{\Lambda}}\boldsymbol{\Psi}_{1}\mathbf{\boldsymbol{\Lambda}}\boldsymbol{\Psi}_{2}\right)\label{def:eta}
\end{equation}
where $\boldsymbol{\Psi}_{1}={\bf U}^{*}\boldsymbol{\Theta}_{1}{\bf U}$
and $\boldsymbol{\Psi}_{2}={\bf V}^{*}\boldsymbol{\Theta}_{2}{\bf V}$.
It is clear that $\boldsymbol{\Psi}_{1}$ and $\boldsymbol{\Psi}_{2}$
are independent matrices that are both distributed as the upper $r\times r$
diagonal block of a Haar unitary matrix.

\section{Result }

The main result of our contribution is the following 
\begin{thm}
\label{thm:main} If $\lambda_{1}(\mathbf{X}_{0})<1$ then 
\begin{align*}
\limsup\mathbb{E}_{0}\left[\mathcal{L}(\mathbf{Y})^{2}\right] & \leq\left(\frac{1}{1-\lambda_{1}(\mathbf{X}_{0})^{4}}\right)^{r^{2}}
\end{align*}
and it is not possible to find a consistent test. 

\begin{comment}
there is no consistent tests for the detection of the spike if $\lambda_{1}(\mathbf{X}_{0})<1.$ 
\end{comment}
\end{thm}
We remind that we are looking for a condition on $\mathbf{X}_{0}$
(due to (\ref{eq:second order moment},\ref{def:eta}), this is a condition
on $\boldsymbol{\Lambda}$) under which $\mathbb{E}\left[\exp\left(2n\eta\right)\right]$
is bounded. Evidently, the divergence may occur only when $\eta>0.$
We hence consider $E_{1}=\mathbb{E}\left[\exp\left(2n\eta\right)\mathbb{I}_{\eta>\epsilon}\right]$
and $E_{2}=\mathbb{E}\left[\exp\left(2n\eta\right)\mathbb{I}_{\eta\leq\epsilon}\right]$,
and prove that, for a certain small enough $\epsilon>0$ to be specified
later, $E_{1}=o(1)$ and that $E_{2}$ is bounded.

\section{The $E_{1}$ term: computation of the GRF of $\eta$.}

It is clear that the boundedness of the integral $E_{1}$ is achieved
when $\eta$ rarely deviates from $0$. As remarked in \cite{montanari-reichman-zeitouni-2017},
the natural machinery to consider is this of the Large Deviation Principle
(LDP). In essence, if $\eta$ follows the LDP with rate $n$, there
can be found a certain non-negative function called Good Rate Function
(GRF) $I_{\eta}$ such that for any Borel set $A$ of $\mathbb{R}$,
$\frac{1}{n}\log\mathbb{P}\left(\eta\in A\right)$ converges towards
$\sup_{x\in A}-I_{\eta}(x)$. The existence of a GRF allows one to
analyze the asymptotic behaviour of the integral $E_{1}$.
In the next section, we thus justify that $\eta$ follows a Large
Deviation Principle with rate $n$, and we compute the associated
GRF.

\subsection{Computation of the GRF of $\eta$}

Eq. (\ref{def:eta}) and the Cauchy-Schwarz inequality imply that the random variable $\eta$ is bounded:
$\left|\eta\right|\leq\eta_{\max}$ with $\eta_{\max}=\sum_{j=1}^{r}\lambda_{j}^{2}.$

We first recall that for $i=1,2$, the random matrix $\boldsymbol{\Psi}_{i}$
follows a LDP with rate $n$ and that its GRF at the parameter $\boldsymbol{\psi}\in\mathbb{C}^{r\times r}$, $\left\Vert \boldsymbol{\psi}\right\Vert _{2}\leq1$, is $\log\det\left(\mathbf{I}_{r}-\boldsymbol{\psi}^{*}\boldsymbol{\psi}\right)$
(see Theorem 3-6 in \cite{gamboa-2014}). Besides, $\eta$ is a function
of the i.i.d. matrices $(\boldsymbol{\Psi}_{i})_{i=1,2}$ and therefore,
the contraction principle applies to $\eta$ (see Theorem 4.2.1 in \cite{dembo-zeitouni2009}):
it ensures that $\eta$ follows a LDP with rate $n$ and its GRF is
such that, for each real $\left|x\right|\leq\eta_{\max}$, $-I_{\eta}(x)$
is the solution of the following optimization problem: 
\begin{problem}
\label{pr:version1} Maximize in $\mathbb{C}^{r\times r}$ 
\begin{align}
\log\det\left(\mathbf{I}-\boldsymbol{\psi}_{1}^{*}\boldsymbol{\psi}_{1}\right)+\log\det\left(\mathbf{I}-\boldsymbol{\psi}_{2}^{*}\boldsymbol{\psi}_{2}\right).\label{eq:optim}
\end{align}
under the constraints 

\begin{eqnarray}
  \mathfrak{R}\mathrm{Tr}\left(\boldsymbol{\Lambda}\boldsymbol{\psi}_{1}\boldsymbol{\Lambda}\boldsymbol{\psi}_{2}\right)=x\label{eq:trace}\\
\|\boldsymbol{\psi}_{i}\|_{\scriptscriptstyle 2}\leq1,\;i=1,2
\end{eqnarray}
\end{problem}
We provide a closed-form solution of Problem \ref{pr:version1}. In
this respect, we define for each $k=1,\ldots,r$ the interval $\mathcal{I}_{k}$
defined by 
\begin{align}
\forall k=1,...,r-1:\ \ \mathcal{I}_{k} & =]\sum_{i=1}^{k}\left(\lambda_{i}^{2}-\lambda_{k}^{2}\right),\sum_{i=1}^{k+1}\left(\lambda_{i}^{2}-\lambda_{k+1}^{2}\right)]\label{eq:interval Ik}
\end{align}
 and $\mathcal{I}_{r}=]\sum_{i=1}^{r}\left(\lambda_{i}^{2}-\lambda_{k}^{2}\right),\eta_{\max}]$.
It is easy to check that $(\mathcal{I}_{k})_{k=1,\ldots,r}$ are disjoint
and that $\cup_{k=1}^{r}\mathcal{I}_{k}=]0,\eta_{\max}]$. The following
result holds: 
\begin{thm}
\label{th:expression-grf} The maximum of Problem \ref{pr:version1}
is given by 
\begin{equation}
-I_{\eta}(x)=2\,\sum_{k=1}^{r}\log\left(\left[\frac{\sum_{i=1}^{k}\lambda_{i}^{2}-|x|}{k}\right]^{k}\frac{1}{\Pi_{i=1}^{k}\lambda_{i}^{2}}\right)\,\mathbb{I}_{\mathcal{I}_{k}}(|x|)\label{eq:expre-grf}
\end{equation}
\end{thm}
It is easy to check that the function $x\mapsto-I_{\eta}(x)$ is continuous
on $]0,\eta_{\max}[$. The proof of Theorem \ref{th:expression-grf}
is provided in the Appendix. 

\begin{figure}
\label{fig:envelope} \includegraphics[scale=0.3]{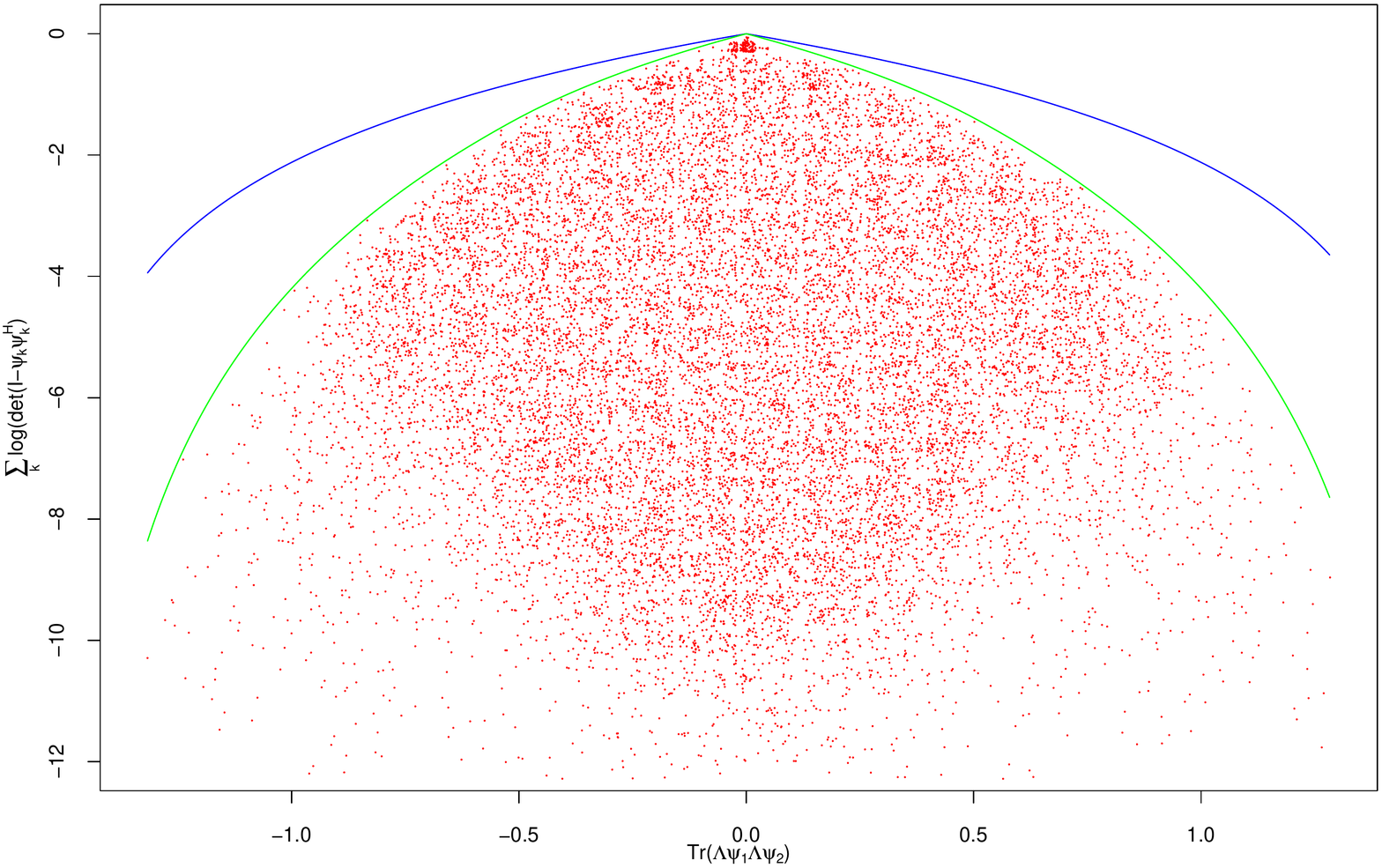}\caption{graph of $-I_{\eta}(x)$ seen as an upper envelope. Upper curve: the
upper bound computed in \cite{Chevreuil:2018db}}
\end{figure}

We illustrate Theorem \ref{th:expression-grf} through the following
experiment. The rank of the spike is fixed to $r=3$ and the singular
values have been set to $\left(\lambda_{1},\lambda_{2},\lambda_{3}\right)=\left(1\ ,\ 0.7\ ,\ 0.2\right)$.
We have computed millions of random samples of the matrices $(\boldsymbol{\psi}_{1},\boldsymbol{\psi}_{2})$.
Each pair is associated with a point $(x,y)$ defined as $x=\mathfrak{R}\mathrm{Tr}\left(\boldsymbol{\Lambda}\boldsymbol{\psi}_{1}\boldsymbol{\Lambda}\boldsymbol{\psi}_{2}\right)$
and $y=\sum_{i=1}^{2}\log\det\left(\mathbf{I}-\boldsymbol{\psi}_{i}^{*}\boldsymbol{\psi}_{i}\right).$
We obtain a cloud of points, the upper envelope of which is expected
to be $-I_{\eta}(x).$ We have also plotted the graph of the function
$y=-I_{\eta}(x)$. In addition, we mention that, in the more general
context of tensors of order $d$, the second-order moment of $\mathcal{L}({\bf Y})$ is still
given by \eqref{eq:second order moment} but the random variable -
call it $\eta_{d}$ - has a more complicated form than (\ref{def:eta}), see
\cite{Chevreuil:2018db};
the asymptotics of the term $E_{1}$ can still be studied by evaluating
the GRF of $\eta_{d}$. This GRF is the solution of an optimization
problem that, apparently,  cannot be  solved in closed form for $d\geq3$.
In \cite{Chevreuil:2018db}, an upper bound of the opposite of the
true GRF is computed; this upper bound, valid for any $d$ is given  for
$d=2$ by $\log\left(1-\frac{\left|x\right|}{\eta_{\max}}\right)$.
We thus also represent in Figure \ref{fig:envelope} this upper bound;
clearly, it is not tight.

\subsection{Computation of $E_{1}$}

The Varadhan lemma (see Theorem 4.3.1 in \cite{dembo-zeitouni2009})
states that $\frac{1}{n}\log\mathbb{E}\left[\exp\left(2n\eta\right)\mathbb{I}_{\eta>\epsilon}\right]\to\sup_{x>\epsilon}\left(2x-I_{\eta}(x)\right)$
and hence the $E_{1}$ term converges towards $0$ when $\sup_{x>\epsilon}\left(2x-I_{\eta}(x)\right)<0$.
Consider any of the intervals $\mathcal{I}_{k}$ defined in \eqref{eq:interval Ik}.
The derivative of $2x-I_\eta(x) $ for any $x\in\mathcal{I}_{k}$
is $2-2k/(\lambda_{1}^{2}+...+\lambda_{k}^{2}-x):$ it is decreasing
on $\mathcal{I}_{k}$ and the limit in the left extremity of $\mathcal{I}_{k}$,
{\it i.e. } $(\sum_{j=1}^{k-1}\lambda_{j}^{2})-(k-1)\lambda_{k}^{2}$, is
simply $2\left(1-\frac{1}{\lambda_{k}^{2}}\right)$. If $\lambda_{1}\left(\mathbf{X}_{0}\right)<1$,
then for all the indices $k$, $1-\frac{1}{\lambda_{k}^{2}}<0$. This
shows that $2x-I_\eta(x)$ is strictly decreasing on every $\mathcal{I}_{k}$.
Hence, for every $x\in]0,\eta_{\max}]$, we have $2x-I_\eta(x)<0-I_\eta (0)=0$.
We have proved that $E_{1}=o(1).$

\section{The $E_{2}$ term: concentration of $\eta$. }

Notice that the upper block $r\times r$ $\boldsymbol{\Psi}$ of a
unitary Haar matrix $\boldsymbol{\Theta}$ has the same distribution
as 
\[
\mathbf{G}\left(\mathbf{\tilde{G}}^{*}\mathbf{\tilde{G}}\right)^{-1/2}
\]
where the $n\times r$ matrix $\tilde{\mathbf{G}}$ has i.i.d. entries
distributed as $\mathcal{N}_{\mathbb{C}}(0,1)$ and $\textbf{G}$
is the top $r\times r$ block of $\tilde{\mathbf{G}}$. Obviously,
$\mathbb{E}[\mathbf{\tilde{G}}^{*}\mathbf{\tilde{G}}]=n\mathbf{I}.$
It is a standard result that a random variable distributed as a $\chi^{2}(n)$
is concentrated around its mean. This can be easily extended to the
matrix $\mathbf{\tilde{G}}^{*}\mathbf{\tilde{G}}$: 
\begin{lem}
For any $0<\delta<1$, there exists a constant $c$ such that 
\[
\mathbb{P}\left(\left\Vert \frac{1}{n}\mathbf{\tilde{G}}^{*}\mathbf{\tilde{G}}-\mathbf{I}\right\Vert_{\scriptscriptstyle 2}>\delta\right)\leq c\exp\left(-n\frac{\delta^{2}}{2}\right).
\]
\end{lem}
We take $\tilde{\mathbf{G}}_{1}$ and $\tilde{\mathbf{G}}_{2}$ independent,
distributed as $\tilde{\mathbf{G}}$ and consider the upper $r\times r$
blocks $\mathbf{G}_{1}$ and $\mathbf{G}_{2}$ of $\tilde{\mathbf{G}}_{1}$
and $\tilde{\mathbf{G}}_{2}$ . It follows that $\eta$ has the same
distribution as $2\mathfrak{R}\text{Tr}\left(\boldsymbol{\Lambda}\mathbf{G}_{1}\left(\mathbf{\tilde{G}}_{1}^{*}\mathbf{\tilde{G}}_{1}\right)^{-1/2}\boldsymbol{\Lambda}\mathbf{G}_{2}\left(\mathbf{\tilde{G}}_{2}^{*}\mathbf{\tilde{G}}_{2}\right)^{-1/2}\right).$
Take now any $\delta<1.$ We may split the integral $E_{2}$ in two
parts: 
\[
\underbrace{\mathbb{E}\left[\exp\left(2n\eta\right)\mathbb{I}_{\left\{ \eta\leq\epsilon\right\} \cap\mathcal{B}_{1}^{c}\cap\mathcal{B}_{2}^{c}}\right]}_{E_{2}'}+\underbrace{\mathbb{E}\left[\exp\left(2n\eta\right)\mathbb{I}_{\left\{ \eta\leq\epsilon\right\} \cap(\mathcal{B}_{1}\cup\mathcal{B}_{2})}\right]}_{E_{2}''}.
\]
where we have defined the events $\mathcal{B}_{i}=\left\{ \left\Vert \frac{1}{n}\mathbf{\tilde{G}}_{i}^{*}\mathbf{\tilde{G}}_{i}-I\right\Vert _{2}>\delta\right\} .$
Thanks to the above concentration result, we have 
\begin{eqnarray*}
E''_{2} & \leq & \exp(2n\epsilon)\left(\mathbb{P}(\mathcal{B}_{1})+\mathbb{P}(\mathcal{B}_{2})\right)\\
 & \leq & \exp(2n\epsilon)2c\exp\left(-n\delta^{2}/2\right)
\end{eqnarray*}

As it is always possible to choose $\delta$ and $\epsilon$ such
that $\delta^{2}-4\epsilon>0$ and $\delta<1$ it follows that $E_{2}''=o(1).$

Let us now inspect the term $E'_{2}.$ Since we have, for $i=1,2$,
$\left\Vert \frac{1}{n}\mathbf{\tilde{G}}_{i}^{*}\mathbf{\tilde{G}}_{i}-I\right\Vert _{2}\leq\delta$,
then there exist $\mathbf{\Delta}_{i}$ for $i=1,2$ such that $\left(\mathbf{\tilde{G}}_{i}^{*}\mathbf{\tilde{G}}_{i}{}\right)^{-1/2}=\frac{1}{\sqrt{n}}\left(\mathbf{I}+\boldsymbol{\Delta}_{i}\right)$with
$\left\Vert \boldsymbol{\Delta}_{i}\right\Vert_{\scriptscriptstyle 2}\leq\delta/2.$
We hence have 
\[
E_{2}'\leq\mathbb{E}\left[\exp\left(2\mathfrak{R}\text{Tr}\left(\boldsymbol{\Lambda}\mathbf{G}_{1}\left(\mathbf{I}+\boldsymbol{\Delta}_{1}\right)\boldsymbol{\Lambda}\mathbf{G}_{2}(\mathbf{I}+\boldsymbol{\Delta}_{2}\right)\right)\right].
\]
We expand $2\mathfrak{R}\text{Tr}\left(\boldsymbol{\Lambda}\mathbf{G}_{1}\left(\mathbf{I}+\boldsymbol{\Delta}_{1}\right)\boldsymbol{\Lambda}\mathbf{G}_{2}\left(\mathbf{I}+\boldsymbol{\Delta}_{2}\right)\right)$
as the sum of four terms. Take for instance 
\[
T_{2}=2\mathfrak{R}\text{Tr}\left(\boldsymbol{\Lambda}\mathbf{G}_{1}\boldsymbol{\Delta}_{1}\boldsymbol{\Lambda}\mathbf{G}_{2}\right)
\]
%\begin{align*}
%T_{1} & =2\mathfrak{R}\text{Tr}\left(\boldsymbol{\Lambda}\mathbf{G}_{1}\boldsymbol{\Lambda}\mathbf{G}_{2}\right)\\
%T_{2} & =2\mathfrak{R}\text{Tr}\left(\boldsymbol{\Lambda}\mathbf{G}_{1}\boldsymbol{\Delta}_{1}\boldsymbol{\Lambda}\mathbf{G}_{2}\right)\\
%T_{3} & =2\mathfrak{R}\text{Tr}\left(\boldsymbol{\Lambda}\mathbf{G}_{1}\boldsymbol{\Lambda}\mathbf{G}_{2}\boldsymbol{\Delta}_{2}\right)\\
%T_{4} & =2\mathfrak{R}\text{Tr}\left(\boldsymbol{\Lambda}\mathbf{G}_{1}\boldsymbol{\Delta}_{1}\boldsymbol{\Lambda}\mathbf{G}_{2}\boldsymbol{\Delta}_{2}\right)
%\end{align*}
Thanks to von Neumann's lemma \cite{VonNeumann}, we have 
\begin{align*}
T_{2} & \leq2\sum_{k=1}^{r}\lambda_{k}(\boldsymbol{\Delta}_{1})\lambda_{k}\left(\boldsymbol{\Lambda}\mathbf{G}_{2}\boldsymbol{\Lambda}\mathbf{G}_{1}\right)\\
 & \leq2\left\Vert \boldsymbol{\Delta}_{1}\right\Vert_{\scriptscriptstyle 2} \sum_{k=1}^{r}\lambda_{k}\left(\boldsymbol{\Lambda}\mathbf{G}_{2}\boldsymbol{\Lambda}\mathbf{G}_{1}\right)
\end{align*}
% where, as usual, $\lambda_{k}()$ denotes the $k$-th singular value
%of a matrix. 
As $\sum_{k=1}^{r}\lambda_{k}\left(\boldsymbol{\Lambda}\mathbf{G}_{2}\boldsymbol{\Lambda}\mathbf{G}_{1}\right)\leq\sqrt{r}\sqrt{\sum_{k=1}^{r}\lambda_{k}^{2}\left(\boldsymbol{\Lambda}\mathbf{G}_{2}\boldsymbol{\Lambda}\mathbf{G}_{1}\right)}$,
it yields 
\begin{align*}
T_{2} & \leq2\left\Vert \boldsymbol{\Delta}_{1}\right\Vert_{\scriptscriptstyle 2} \sqrt{r}\sqrt{\text{Tr}\left(\boldsymbol{\Lambda}\mathbf{G}_{2}\boldsymbol{\Lambda}\mathbf{G}_{1}\mathbf{G}_{1}^{*}\boldsymbol{\Lambda}\mathbf{G}_{2}^{*}\boldsymbol{\Lambda}\right)}.
\end{align*}
%Invoking again von Neumann's lemma gives finally  $\text{Tr}\left(\left(\mathbf{G}_{2}^{*}\boldsymbol{\Lambda}^{2}\mathbf{G}_{2}\right)\left(\boldsymbol{\Lambda}\mathbf{G}_{1}\mathbf{G}_{1}^{*}\boldsymbol{\Lambda}\right)\right)\leq\sum_{k}\lambda_{k}\left(\mathbf{G}_{2}^{*}\boldsymbol{\Lambda}^{2}\mathbf{G}_{2}\right)\lambda_{k}\left(\boldsymbol{\Lambda}\mathbf{G}_{1}\mathbf{G}_{1}^{*}\boldsymbol{\Lambda}\right)$,
%thus implying 
%\[
%\text{Tr}\left(\boldsymbol{\Lambda}\mathbf{G}_{2}\boldsymbol{\Lambda}\mathbf{G}_{1}\mathbf{G}_{1}^{*}\boldsymbol{\Lambda}\mathbf{G}_{2}^{*}\boldsymbol{\Lambda}\right)\leq\underbrace{\sum_{k}\lambda_{k}\left(\mathbf{G}_{2}^{*}\boldsymbol{\Lambda}^{2}\mathbf{G}_{2}\right)}_{\text{Tr }\left(\mathbf{G}_{2}^{*}\boldsymbol{\Lambda}^{2}\mathbf{G}_{2}\right)}\underbrace{\sum_{k}\lambda_{k}\left(\boldsymbol{\Lambda}\mathbf{G}_{1}\mathbf{G}_{1}^{*}\boldsymbol{\Lambda}\right)}_{\text{Tr }\left(\boldsymbol{\Lambda}\mathbf{G}_{1}\mathbf{G}_{1}^{*}\boldsymbol{\Lambda}\right)}.
%\]
%In the same manner, $\text{Tr }\left(\mathbf{G}_{2}^{*}\boldsymbol{\Lambda}^{2}\mathbf{G}_{2}\right)\leq\left\Vert \boldsymbol{\Lambda}^{2}\right\Vert \text{Tr }\left(\mathbf{G}_{2}^{*}\mathbf{G}_{2}\right)$
%and $\text{Tr }\left(\boldsymbol{\Lambda}\mathbf{G}_{1}\mathbf{G}_{1}^{*}\boldsymbol{\Lambda}\right)\leq\left\Vert \boldsymbol{\Lambda}^{2}\right\Vert \text{Tr }\left(\mathbf{G}_{1}\mathbf{G}_{1}^{*}\right).$ 

Invoking the von Neumann's lemma three times, it holds that 
\begin{align*}
T_{2} & \leq2\left\Vert \boldsymbol{\Delta}_{1}\right\Vert_{\scriptscriptstyle 2} \sqrt{r}\left\Vert \boldsymbol{\Lambda}^{2}\right\Vert_{\scriptscriptstyle 2} \sqrt{\text{Tr }\left(\mathbf{G}_{1}\mathbf{G}_{1}^{*}\right)\text{Tr }\left(\mathbf{G}_{2}\mathbf{G}_{2}^{*}\right)}\\
 & \leq\sqrt{r}\left\Vert \boldsymbol{\Delta}_{1}\right\Vert_{\scriptscriptstyle 2} \left\Vert \boldsymbol{\Lambda}^{2}\right\Vert_{\scriptscriptstyle 2} \left(\text{Tr }\left(\mathbf{G}_{1}\mathbf{G}_{1}^{*}\right)+\text{Tr }\left(\mathbf{G}_{2}\mathbf{G}_{2}^{*}\right)\right)
\end{align*}
Similar manipulations can be done on the other terms of the expansion.
%We can show in the same spirit that 
%\begin{align*}
%T_{3} & \leq\sqrt{r}\left\Vert \boldsymbol{\Delta}_{2}\right\Vert \left\Vert \boldsymbol{\Lambda}\right\Vert ^{2}\left(\text{Tr }\left(\mathbf{G}_{1}\mathbf{G}_{1}^{*}\right)+\text{Tr }\left(\mathbf{G}_{2}\mathbf{G}_{2}^{*}\right)\right)\\
%T_{4} & \leq\sqrt{r}\left\Vert \boldsymbol{\Delta}_{1}\right\Vert \left\Vert \boldsymbol{\Delta}_{2}\right\Vert \left\Vert \boldsymbol{\Lambda}\right\Vert ^{2}\left(\text{Tr }\left(\mathbf{G}_{1}\mathbf{G}_{1}^{*}\right)+\text{Tr }\left(\mathbf{G}_{2}\mathbf{G}_{2}^{*}\right)\right)
%\end{align*}
so that $E_{2}'$ is less than
\[
\mathbb{E}\left[\exp\left(2\mathfrak{R}\text{Tr}\left(\boldsymbol{\Lambda}\mathbf{G}_{1}\boldsymbol{\Lambda}\mathbf{G}_{2}\right)+\beta\text{Tr}\left(\left(\mathbf{G}_{1}\mathbf{G}_{1}^{*}\right)+\text{Tr}\left(\mathbf{G}_{2}\mathbf{G}_{2}^{*}\right)\right)\right)\right]
\]
with $\beta=\frac{\sqrt{r}}{2}\delta(2+\delta)\left\Vert \boldsymbol{\Lambda}\right\Vert_{\scriptscriptstyle 2} ^{2}.$
The above expectation is to be understood as the expectation over $(\mathbf{G}_{1}, \mathbf{G}_{2})$. As $\mathbf{G}_{1}$ and $\mathbf{G}_{2}$ are independent, we consider
first the expectation over $\mathbf{G}_{1}$. This gives, up to the
factor $\exp\left(\beta\text{Tr }\left(\mathbf{G}_{2}\mathbf{G}_{2}^{*}\right)\right)$
\[
\pi^{-r^{2}}\int\exp\left(2\mathfrak{R}\text{Tr}\left(\mathbf{g}_{1}\mathbf{E}\right)+\left(\beta-1\right)\text{Tr }\left(\mathbf{g}_{1}^{*}\mathbf{g}_{1}\right)\ \right)d\mathbf{g}_{1}
\]
with $\mathbf{E}=\boldsymbol{\Lambda}\mathbf{G}_{2}\boldsymbol{\Lambda}.$
It is always possible to choose $\delta$ such that $\beta<1$. With
such a $\beta$, the above integral is %\begin{align*}
%\left(1-\beta\right)^{-r^{2}}\pi^{-r^{2}}\int\exp\left(\frac{2}{\sqrt{1-\beta}}\mathfrak{R}\text{Tr}\left(\mathbf{g}_{1}\mathbf{E}\right)-\text{Tr }\left(\mathbf{g}_{1}^{*}\mathbf{g}_{1}\right)\ \right)d\mathbf{g}_{1}
%\end{align*}
\[
\left(1-\beta\right)^{-r^{2}}\exp\left(\frac{1}{4}\left(\frac{2}{\sqrt{1-\beta}}\right)^{2}\text{Tr}\left(\mathbf{E}\mathbf{E}^{*}\right)\right)
\]
As $\text{Tr}\left(\mathbf{E}\mathbf{E}^{*}\right)\leq\left\Vert \boldsymbol{\Lambda}\right\Vert _{2}^{4}\text{Tr}\left(\mathbf{G}_{2}\mathbf{G}_{2}^{*}\right)$
we finally obtain after multiplying by $\exp\left(\beta\text{Tr }\left(\mathbf{G}_{2}\mathbf{G}_{2}^{*}\right)\right)$
and taking the expectation over $\mathbf{G}_{2}$,  $E_{2}'$ is less or equal to
\[
\frac{\left(1-\beta\right)^{-r^{2}}}{\pi^{r^{2}}}\int\exp\left(-\frac{(1-\beta)^{2}-\left\Vert \Lambda\right\Vert _{2}^{4}}{1-\beta}\text{Tr}\left(\mathbf{g}_{2}^{*}\mathbf{g}_{2}\right)\right)d\mathbf{g}_{2}.
\]
If $\left\Vert \boldsymbol{\Lambda}\right\Vert _{2}^{2}<1$, it is
always possible to adjust $\delta$ such that the above integral converges.
In this condition, we have 
\[
E_{2}'\leq\left(\frac{1}{(1-\beta)^{2}-\left\Vert \boldsymbol{\Lambda}\right\Vert _{2}^{4}}\right)^{r^{2}}.
\]
This must be true for all $\beta$ arbitrarily small, hence the result.

\appendix

 We prove Theorem \ref{th:expression-grf} when $x>0$. As the function to be maximized converges towards
$-\infty$ if $\|\boldsymbol{\psi}_{1}\|\rightarrow1$ or $\|\boldsymbol{\psi}_{2}\|\rightarrow1$,
any argument $(\boldsymbol{\psi}_{1},\boldsymbol{\psi}_{2})$ of the
maximization problem satisfies $\|\boldsymbol{\psi}_{i}\|<1,\;i=1,2$.
Therefore, the Karush-Kuhn-Tucker (KKT) conditions imply the existence
of a scalar Lagrange multiplier $\mu\geq0$ such that $(\boldsymbol{\psi}_{1},\boldsymbol{\psi}_{2})$
is a stationary point of the Lagrangian $\ell(\boldsymbol{\psi}_{1},\boldsymbol{\psi}_{2},\mu)$
defined by $\sum_{i=1}^{2}\log\det\left(\mathbf{I}_{r}-\boldsymbol{\psi}_{i}^{*}\boldsymbol{\psi}_{i}\right)+\mu\;\mathfrak{R}\mathrm{Tr}\left(\boldsymbol{\Lambda}\boldsymbol{\psi}_{1}\boldsymbol{\Lambda}\boldsymbol{\psi}_{2}\right).$ As $\ell$ is a real valued function, a stationary point is computed when setting the differential w.r.t. the entries of $\boldsymbol{\psi}_{1}$
and $\boldsymbol{\psi}_{2}$ to zero. It can be checked that $(\boldsymbol{\psi}_{1},\boldsymbol{\psi}_{2})$
is a stationary point of $\ell$ when

\begin{align*}
\mu\,\boldsymbol{\Lambda}\boldsymbol{\psi}_{2}\boldsymbol{\Lambda} & =\boldsymbol{\psi}_{1}^{*}({\bf I}-\boldsymbol{\psi}_{1}\boldsymbol{\psi}_{1}^{*})^{-1}\\
\mu\,\boldsymbol{\Lambda}\boldsymbol{\psi}_{1}\boldsymbol{\Lambda} & =\boldsymbol{\psi}_{2}^{*}({\bf I}-\boldsymbol{\psi}_{2}\boldsymbol{\psi}_{2}^{*})^{-1}
\end{align*}

In a first step, these equations can be shown to be satisfied only
if $\boldsymbol{\psi}_{1}$ and $\boldsymbol{\psi}_{2}$ are diagonal
up to permutations of the columns. Then, is can be deduced that there
exists a diagonal matrix $0\leq\mathbf{P}\leq\mathbf{I}$ and a matrix
of permutation $\boldsymbol{\Pi}$ such that $\log\det\left(\mathbf{I}_{r}-\boldsymbol{\psi}_{1}^{*}\boldsymbol{\psi}_{1}\right)+\log\det\left(\mathbf{I}_{r}-\boldsymbol{\psi}_{2}^{*}\boldsymbol{\psi}_{2}\right)=2\log\mathrm{det}({\bf I}-{\bf P})$and
$\mathfrak{R}\mathrm{Tr}\left(\boldsymbol{\Lambda}\boldsymbol{\psi}_{1}\boldsymbol{\Lambda}\boldsymbol{\psi}_{2}\right)=\mathrm{Tr}(\boldsymbol{\Lambda}\boldsymbol{\Pi}^{*}\boldsymbol{\Lambda}\boldsymbol{\Pi}\,{\bf P})$.
This invites us to consider the following 
\begin{problem}
\label{pr:problem-with-permutation}Maximize 
\begin{equation}
\log\mathrm{det}({\bf I}-{\bf P})\label{eq:reduced-optimization-problem}
\end{equation}
jointly over all the $r!$ permutations $\boldsymbol{\Pi}$ and over
diagonal matrices ${\bf P}$ verifying $0\leq{\bf P}\leq{\bf I}$
and the constraint 
\begin{equation}
\mathrm{Tr}(\boldsymbol{\Lambda}\boldsymbol{\Pi}^{*}\boldsymbol{\Lambda}\boldsymbol{\Pi}\,{\bf P})=x.\label{eq:new-constraint}
\end{equation}
\end{problem}
In a first step, we set $\boldsymbol{\Pi}=\mathbf{I}$ in the above
problem and consider the 
\begin{problem}
\label{pr:reduced-optimization-problem} Maximize 
\begin{equation}
\sum_{i=1}^{r}\log(1-p_{i})\label{eq:reduced-optimization-problem-I}
\end{equation}
under the constraints that $0\leq p_{i}\leq1$ for each $i=1,\ldots,r$
and 
\begin{equation}
\sum_{i=1}^{r}\lambda_{i}^{2}\,p_{i}=x.\label{eq:new-constraint-I}
\end{equation}
The maximum is denoted by $J_{\Lambda}(x).$ 
\end{problem}
This is a variant of the celebrated water-filling problem (see {\it e.g. } 
\cite{witsenhausen-1975} and Chap. 9 of \cite{cover-thomas}) that
was solved to evaluate the capacity of a frequency selective Gaussian
channel, the difference being that in the latter problem, $\log(1-p_{i})$
is replaced by $\log(1+p_{i})$. %  while the constraint $p_{i}\leq1$ disappears. 
  In order to solve Problem \ref{pr:reduced-optimization-problem},
we assume that the non zero singular values $(\lambda_{i})_{i=1,\ldots,r}$
are distinct. If this is not the case, a standard perturbation argument
can be used in order to address the general case. As the function
to be maximized is strictly concave on the set defined by the constraints,
the maximum  is reached at a unique point ${\bf p}_{*}$ verifying
$p_{i,*}<1$ for each $i$. We consider the Lagrangian corresponding
to Problem (\ref{pr:reduced-optimization-problem}) given by $\sum_{i=1}^{r}\log(1-p_{i})+\mu\left(\sum_{i=1}^{r}\lambda_{i}^{2}\,p_{i}\right)+\sum_{i=1}^{r}\delta_{i}p_{i}$
where $\mu\geq0$ and $\delta_{i}\geq0$ for $i=1,\ldots,r$. The
partial derivatives w.r.t. parameters $(p_{i})_{i=1,\ldots,r}$ are
zero at ${\bf p}_{*}$. This leads to 
\begin{equation}
\text{for }i=1,\ldots,r:\ \ \frac{1}{1-p_{i,*}}=\mu_{*}\lambda_{i}^{2}+\delta_{i,*}\label{eq:lagrange-delta}
\end{equation}
The first remark is that necessarily, these equations imply that the
numbers $p_{i,*}$ are sorted in decreasing order. To verify this
claim, we assume that $i<j$ and that $p_{i,*}=0$ and $p_{j,*}>0$.
Then, it holds that $\mu_{*}\lambda_{i}^{2}+\delta_{i,*}=1$ and that
$\mu_{*}\lambda_{j}^{2}=\frac{1}{1-p_{j,*}}>1$ because $p_{j,*} > 0$ implies $\delta_{j,*} = 0$. Therefore, $\lambda_{i}^{2}\leq\frac{1}{\mu_{*}}<\lambda_{j}^{2}$,
a contradiction because $\lambda_{i}^{2}\geq\lambda_{j}^{2}$. We
denote by $s(x)$ the number of non-zero entries of ${\bf p}_{*}$.
Hence, the first $s(x)$ entries of ${\bf p}_{*}$ are non zero. Morever,
the equations $\mu_{*}\lambda_{i}^{2}=\frac{1}{1-p_{i,*}}$ for $i=1,\ldots,s(x)$
imply that $p_{1,*}\geq\ldots\geq p_{s(x),*}>0=p_{s(x)+1,*}=\ldots=p_{r,*}$.

We now analytically characterize $s(x)$. On the one hand, (\ref{eq:lagrange-delta})
computed at for $i=s(x)$ and for $i=s(x)+1$
both imply 
\begin{equation}
\lambda_{s(x)+1}^{2}\leq\frac{1}{\mu_{*}}<\lambda_{s(x)}^{2}\label{eq:encadrement 1/mu}
\end{equation}
On the other hand, the constraint \eqref{eq:new-constraint-I} imposes
that $1/\mu_{*}$ verifies 
\[
\frac{1}{\mu_{*}}=\frac{\sum_{i=1}^{s(x)}\lambda_{i}^{2}-x}{s(x)}.
\]
Therefore, it holds that 
\begin{equation}
(\sum_{i=1}^{s(x)}\lambda_{i}^{2})-s(x)\lambda_{s(x)}^{2}<x\leq(\sum_{i=1}^{s(x)}\lambda_{i}^{2})-s(x)\lambda_{s(x)+1}^{2}\label{eq:encadrement x}
\end{equation}
such that $s(x)$ coincides with the integer $k$ for which $x\in\mathcal{I}_{k}$
(see \eqref{eq:interval Ik} for the definition of these intervals).
The maximum $\sum_{i=1}^{s(x)}\log(1-p_{i,*})$ is direcly computed
as 
\begin{equation}
J_{\Lambda}(x)=\log\left(\left[\frac{\sum_{i=1}^{s(x)}\lambda_{i}^{2}-x}{s(x)}\right]^{s(x)}\frac{1}{\Pi_{i=1}^{s(x)}\lambda_{i}^{2}}\right)\label{eq:expression-maximum}
\end{equation}

In order to show that the GRF of $\eta$ is $I_{\eta}(x)=-2J_{\Lambda}(x),$
it remains to show that the solution of Problem \ref{pr:problem-with-permutation}
is reached when the permutation matrix $\boldsymbol{\Pi}$ is the
identity. In this respect, we introduce a nested problem motivated
by the following observation. We denote by $\boldsymbol{\alpha}$
and $\boldsymbol{\beta}$ the $r$\textendash dimensional vectors
whose components are respectively the diagonal entries of $\boldsymbol{\Lambda}^{2}$
and of $\boldsymbol{\Lambda}\boldsymbol{\Pi}^{*}\boldsymbol{\Lambda}\boldsymbol{\Pi}$
arranged in the decreasing order. Evidently, $\boldsymbol{\alpha}$
majorizes $\boldsymbol{\beta}$ in the sense that 
\begin{equation}
\text{for }k=1,\ldots,r:\ \ \sum_{i=1}^{k}\alpha_{i}\geq\sum_{i=1}^{k}\beta_{i}\label{eq:def-majorization}
\end{equation}
We thus consider the relaxed problem 
\begin{problem}
\label{pr:relaxed} Maximize $\log\mathrm{det}({\bf I}-{\bf P})$
over the diagonal matrices $0\leq{\bf P}\leq{\bf I}$ and over vectors
$\boldsymbol{\beta}=(\beta_{1},...,\beta_{r})$ satisfying $\beta_{1}\geq\beta_{2}\geq\ldots\beta_{r}\geq0$,
the majorization constraint (\ref{eq:def-majorization}), and the
equality constraint 
\begin{equation}
\sum_{i=1}^{r}\beta_{i}\,p_{i}=x\label{eq:constraints-beta-1}
\end{equation}
\end{problem}
The maximum of Problem \ref{pr:relaxed} is above
 the maximum of Problem \ref{pr:problem-with-permutation} which
is itself above  the maximum $J_{\Lambda}(x)$ of Problem \ref{pr:reduced-optimization-problem}.
We actually show that the maximum of Problem \ref{pr:relaxed} is
less than $J_{\Lambda}(x)$, and that it is reached for a vector $\boldsymbol{\beta}$
that coincides with $\boldsymbol{\alpha}$. This will imply that the
optimal permutation $\boldsymbol{\Pi}$ in Problem \ref{pr:problem-with-permutation}
is $\mathbf{I}$ and  $I_{\eta}(x)=-2J_{\Lambda}(x)$.

We give some elements for solving Problem \ref{pr:relaxed}. We consider
a stationary point $\left(\mathbf{p}_{*},\boldsymbol{\beta}_{*}\right)$
of the associated Lagrangian and compute the KKT conditions. We suppose
that this stationary point attains the maximum. If $s$ denotes the
number of non-zero components in $\mathbf{p}_{*}$, we prove that, necessarily, $p_{1,*}\geq p_{2,*}\geq...\geq p_{s,*}>0$
and $\beta_{1,*}\geq\beta_{2,*}\geq...\geq\beta_{s,*}.$ We let $j_{1}$
be the first index such that $\sum_{i=1}^{j_{1}}\alpha_{i}>\sum_{i=1}^{j_{1}}\beta_{i}$
(this index exists otherwise $\boldsymbol{\beta}_{*}=\boldsymbol{\alpha}$
and the problem is solved). This implies that $\beta_{i,*}=\alpha_{i}$
for all indices $i=1,...,j_{1}-1$. Notice this fact: if we suppose
that the condition $\sum_{i=1}^{j_{1}+k}\alpha_{i}>\sum_{i=1}^{j+k}\beta_{i}$
is true whatever $k$, then it is possible to add a small $\epsilon>0$,
and update $\beta_{j_{1,*}}$ as $\beta_{j_{1},*}+\epsilon$ in such
a way that the majorization constraints still hold, the constraint
\eqref{eq:constraints-beta-1} holds and the updated $\mathbf{p_{*}}$
increases the function to maximize. This is in contradiction with
the definition of $\left(\mathbf{p}_{*},\boldsymbol{\beta}_{*}\right)$.
This means that there exists an index $j_{2}$ (we choose the smallest)
such that $\sum_{i=1}^{j_{1}+j_{2}}\alpha_{i}=\sum_{i=1}^{j_{1}+j_{2}}\beta_{i,*}.$
It can be shown that it is necessary that all the $\beta_{i,*}$ are
equal for $i=j_{1},...,j_{1}+j_{2}.$ After some algebraic gymnastics,
it can be shown that it in this case, all the inequalities \eqref{eq:def-majorization}
at $\boldsymbol{\beta}_{*}$ are saturated hence implying that $\boldsymbol{\beta}_{*}=\boldsymbol{\alpha}.$
The value of $\sum_{i}\log(1-p_{i,*})$ equals $J_{\Lambda}(x)$.

 \bibliographystyle{plain}
\bibliography{biblioEUSIPCO18}

\end{document}